\documentclass[twocolumn,pra,unsortedaddress,superscriptaddress,floatfix,citeautoscript,nofootinbib,longbibliography]{revtex4-1}
\usepackage{adjustbox}
\usepackage{dcolumn}% Align table columns on decimal point
\usepackage{latexsym,epsfig,graphicx,dcolumn,subfigure,comment,ulem}
\usepackage{amsmath,eqnarray,amssymb,amsbsy}
\usepackage{xcolor,color}
\usepackage[colorlinks,urlcolor=blue,citecolor=blue]{hyperref}
\usepackage{mathrsfs}

\begin{document}

\title{Edge supercurrent in Josephson junctions based on topological materials}

\author{Junjie Qi}
\email{qijj@baqis.ac.cn}
\affiliation{Beijing Academy of Quantum Information Sciences, 
	Beijing 100193, China}
 
 \author{Chui-Zhen Chen}
 \affiliation{School of Physical Science and Technology, Soochow University, Suzhou 215006, China}
\affiliation{
Institute for Advanced Study, Soochow University, Suzhou 215006, China}

\author{Juntao Song}
\affiliation{College of Physics and Hebei Advanced Thin Films Laboratory, Hebei Normal University, Shijiazhuang, Hebei 050024, China}

\author{Jie Liu}
\email{jieliuphy@xjtu.edu.cn}
\affiliation{School of Physics, Xi’an Jiaotong University, Ministry of Education Key Laboratory for Non-Equilibrium Synthesis
and Modulation of Condensed Matter, Xi’an 710049, China}

\author{Ke He}
\affiliation{State Key Laboratory of Low Dimensional Quantum Physics, Department of Physics, Tsinghua University, Beijing 100084, China}
\affiliation{Beijing Academy of Quantum Information Sciences, 
	Beijing 100193, China}
\affiliation{Frontier Science Center for Quantum Information, Beijing 100084, China}
\affiliation{Hefei National Laboratory, Hefei 230088, China}

\author{Qing-Feng Sun}
\affiliation{International Center for Quantum Materials, School of Physics, Peking University, Beijing 100871, China}
\affiliation{Hefei National Laboratory, Hefei 230088, China}

\author{X. C. Xie}
%\email{xcxie@pku.edu.cn}
\affiliation{International Center for Quantum Materials, School of Physics, Peking University, Beijing 100871, China}
\affiliation{
Interdisciplinary Center for Theoretical Physcics and Information Sciences, Fudan University, Shanghai 200433, China}
\affiliation{Hefei National Laboratory, Hefei 230088, China}

\begin{abstract}

The interplay between novel topological states and superconductivity has garnered substantial interest due to its potential for topological quantum computing. The Josephson effect serves as a useful probe for edge superconductivity in these hybrid topological materials. In Josephson junctions based on topological materials,  supercurrents exhibit unique quantum interference patterns, including the conventional Fraunhofer oscillations, the $\Phi_0$-periodic oscillation, and the $2\Phi_0$-periodic oscillation in response to the external magnetic field ($\Phi_0 = h/2e$ is the flux quantum, $h$ the Planck constant, and $e$ the electron charge). These interference patterns stem from varied Andreev reflection mechanisms and the associated current density profiles.  This review seeks to comprehensively examine the theoretical and experimental advancements in understanding the quantum interference patterns of edge supercurrents in Josephson junctions based on quantum spin Hall, quantum Hall, and quantum anomalous Hall systems.

\end{abstract}

\maketitle
\section{Introduction}\label{section1}

\begin{figure*}[ht!]
    \centering
    \includegraphics[width=1\linewidth]{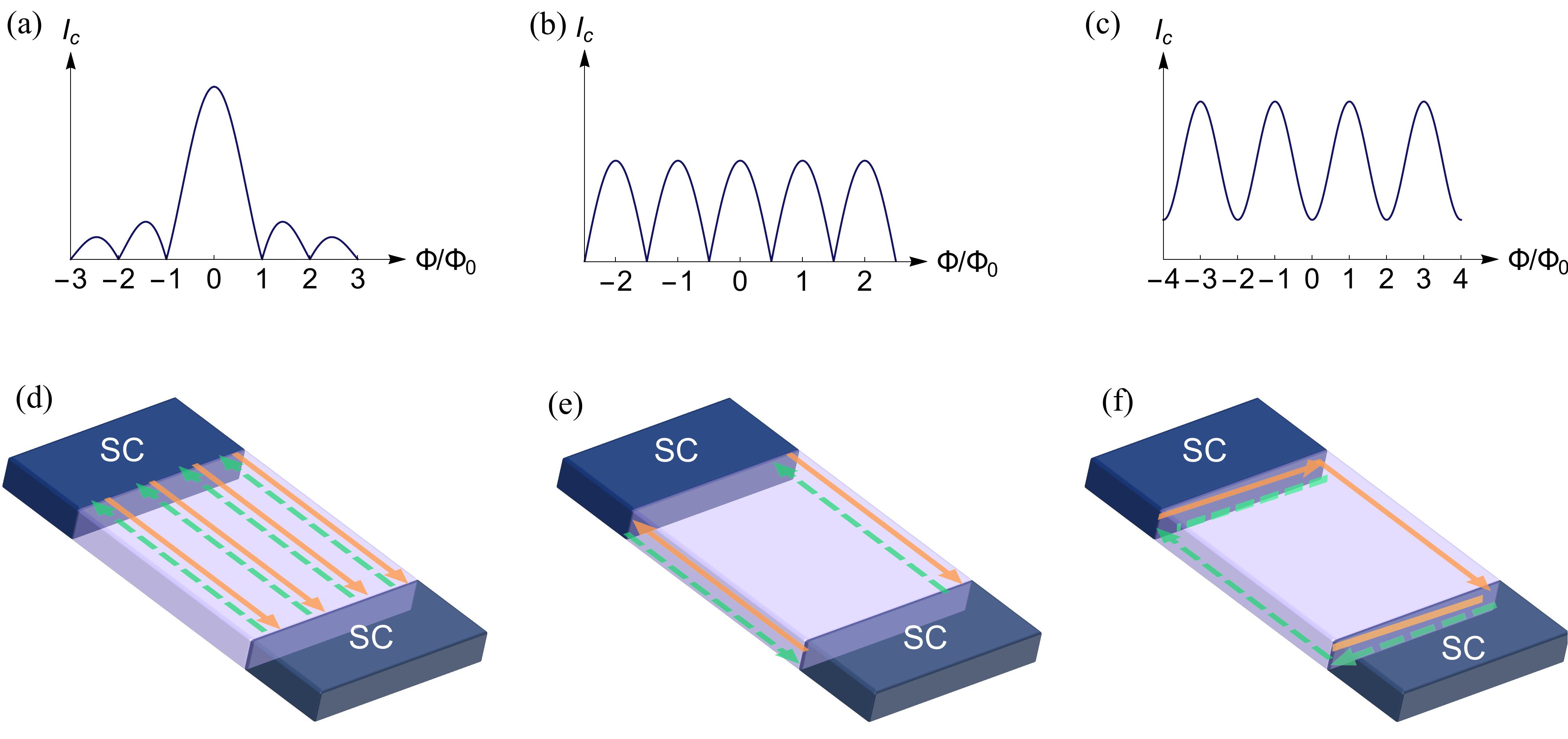}
    \caption{(a) When the central region is filled with bulk carriers, LARs uniformly occur at the interfaces between the SC and the central region. Consequently, the Josephson critical current  $I_c$ oscillates with the external magnetic field, resulting in a Fraunhofer pattern. (b) In the QSH-based JJs, LARs occur at the single helical edge states. The resulting edge-dominated supercurrents exhibit a $\Phi_0$-periodic oscillation. (c) In the QH-based and QAH-based JJs, edge supercurrents arise from CARs instead of LARs. This leads to a $2\Phi_0$-periodic oscillation. The schematic diagrams illustrate (d) LARs in the conventional JJs, (e) LARs at the helical edge states in the QSH-based JJs, and (f) CARs at the chiral edge states in the QH-based and QAH-based JJs. The orange (green) lines represent the electrons (holes), respectively. }
    \label{fig:fig1}
\end{figure*}

The research on topological materials is at the forefront of condensed matter physics, contributing both to fundamental science and potential technological breakthroughs. The quantum Hall effect (QHE), first discovered by Klitzing  \cite{ref1,ref2,ref3} in 1980, exemplifies the manifestation of topological phenomena in condensed matter physics. 
 The QHE is distinguished by chiral edge states, where electrons propagate unidirectionally along the sample's edges while the bulk remains insulating. This effect occurs in a two-dimensional electron gas under a strong magnetic field at low temperatures, leading to quantized Hall resistance plateaus at 
{\color{black} $h/ne^2$ ($n$ is an integer)}, where $h$ is Planck’s constant and $e$ is the elementary charge \cite{ref1,ref4}. The quantized Hall resistance can be understood through topological invariants known as Chern or Thouless–Kohmoto–Nightingale–Nijs (TKNN) numbers \cite{ref4,CN2}. The quantum anomalous Hall effect (QAHE) represents a variant of the QHE that does not require an external magnetic field, yet still supports chiral edge states \cite{ref5,ref6}. The QAHE was theoretically predicted by Haldane \cite{ref7} in 1988 for a two-dimensional honeycomb lattice without a magnetic field  and experimentally observed in Cr-doped $(\mathrm{Bi}, \mathrm{Sb})_2 \mathrm{Te}_3$ thin films in 2013 \cite{ref8}. Since then, QAHE has been observed in various materials, including 
Cr-doped $\rm{(Bi,Sb)_2Te_3}$ films \cite{QAH11,QAH12,QAH13,QAH14,QAH15,QAH16,QAH17,QAH18,QAH19}, V-doped  $\rm{(Bi,Sb)_2Te_3}$ films \cite{QAH4,QAH12,QAH21,QAH22,QAH23,QAH24}, $\rm{MnBi_2Te_4}$ \cite{QAH5,QAH31,QAH32}, and recently moiré superlattice systems \cite{QAH6,QAH7,QAH8}. Both QHE and QAHE break time-reversal symmetry, whereas the quantum spin Hall effect (QSHE) preserves it. The QSHE is characterized by metallic helical edge states protected by a bulk gap \cite{QSH1,QSH2,QSH11,QSH22,QSH3,QSH4}. The helical edge states mean that electrons with opposite spins counter-propagate along the edges of the material. This can be regarded as two copies of QAHE with opposite Chern numbers. Experimentally, QSHE has been observed in HgTe/CdTe quantum wells \cite{QSH5},  InAs/GaSb double quantum wells \cite{QSH6}, and   monolayer $\rm{WTe}_2$ \cite{QSH7,QSH8}.

Remarkably, the interplay between superconductivity and QSHE/QHE/QAHE has attracted considerable attention for its potential to realize topological superconductors (TSCs) capable of hosting Majorana fermions \cite{MF1,QSH22,FJE1}. Majorana fermions, which are exotic quasiparticles that are their own antiparticles, exhibit non-Abelian statistics and are promising candidates for topological quantum computing \cite{TQC1,TQC2,TQC3,TQC4,TQC5,TQC6,TQC7,TQC8,TQC9}. Their nonlocal nature confers resistance to local perturbations, providing a robust platform for fault-tolerant quantum computation. Consequently, the study of TSCs is not only of fundamental importance in condensed matter physics but also essential for the advancement of next-generation quantum technologies. Establishing the presence of edge superconductivity is a critical step toward demonstrating topological superconductivity in these systems, and the Josephson effect serves as a key tool for probing this edge superconductivity.

The Josephson effect, discovered by Josephson in 1962 \cite{JJ1}, arises when two superconductors (SC) are separated by a central region that can be an insulator, normal metal, semiconductor, {\color{black} or a narrow superconductor}. This configuration, known as a Josephson junction (JJ), permits the tunneling of Cooper pairs across the central region. The hallmark of the Josephson effect is the Josephson current, a supercurrent flowing through the junction driven by the superconducting phase difference without any applied voltage, known as the DC Josephson effect.  The generation of this supercurrent is facilitated by Andreev reflections, where an electron converts into a hole at the interfaces between the SC and central region, concurrently creating a Cooper pair in the superconducting region \cite{AR}. In the conventional SC-normal metal-SC (SNS) JJs, local Andreev reflections (LARs) occur uniformly within the central region (see Fig.~\ref{fig:fig1}(d)), which means an electron is retroreflected into a hole, leading to bulk supercurrent oscillations in response to an external magnetic field. As shown in Fig.~\ref{fig:fig1}(a), these oscillations produce a quantum interference pattern known as the Fraunhofer pattern. The Fraunhofer pattern is characterized by a central lobe with a width of $2\Phi_0$ and decaying side lobes. Here, $\Phi_0 = h/2e$ is the flux quantum, $h$ the Planck constant, and $e$ the electron charge. Recently, the combination of the JJs and topological materials has garnered significant attention due to the novel quantum phenomena and potential applications in quantum computing. When the central region exhibits QSHE, the junction is referred to as a QSH-based JJ. Unlike traditional JJs, bulk supercurrent is prohibited due to the presence of a bulk gap. Instead, edge supercurrent, carried by helical edge states, can also sustain even when the chemical potential resides within the bulk gap. In this regime, LARs can occur at a single helical edge (see Fig.~\ref{fig:fig1}(e)). Thus, the edge supercurrent oscillates with the magnetic field resulting in a ``SQUID-like" quantum interference pattern characterized by a $\Phi_0$ periodicity depicted in Fig.~\ref{fig:fig1}(b) . 

\begin{figure*}[ht!]
    \centering
    \includegraphics[width=0.8\linewidth]{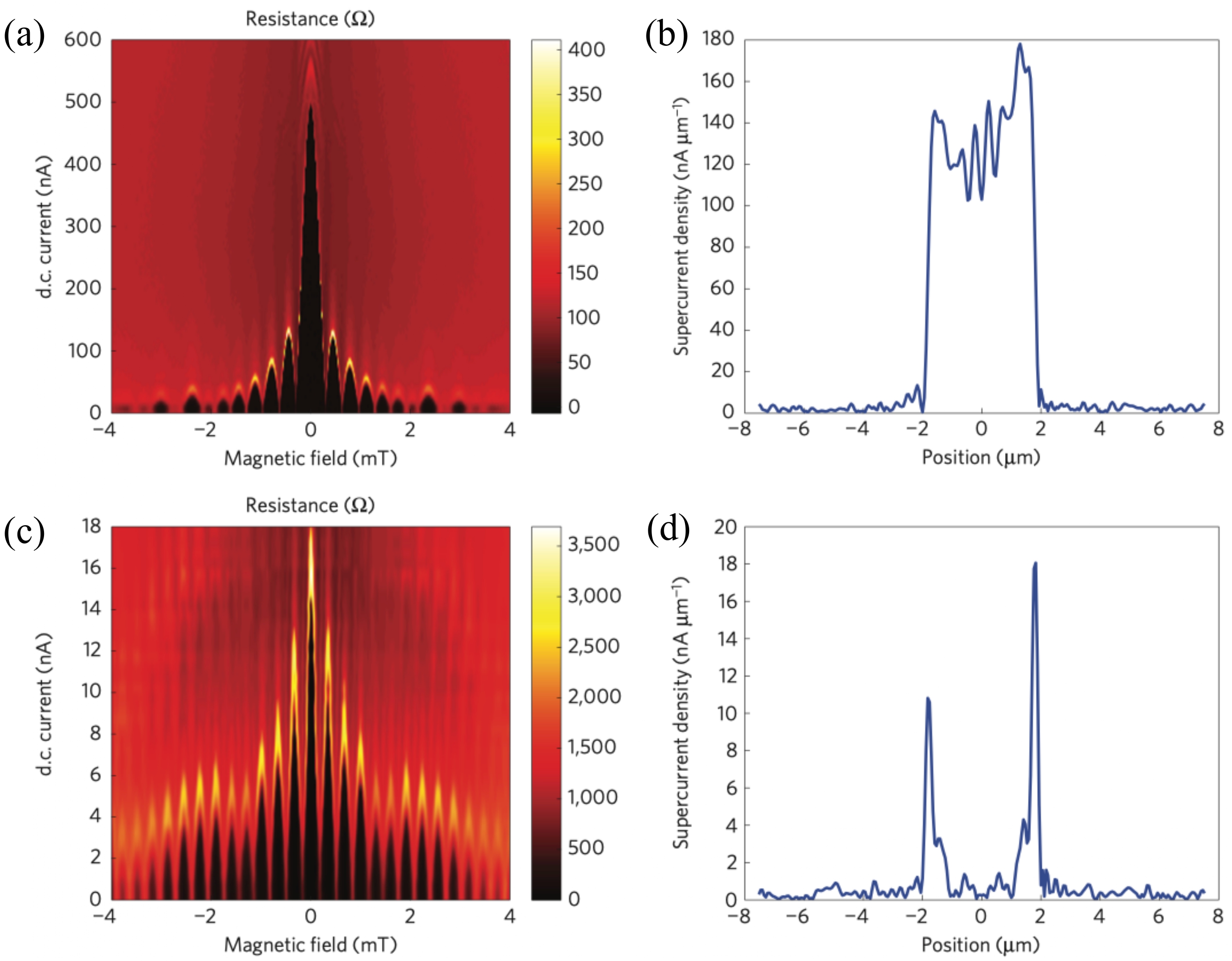}
    \caption{(a) Differential resistance (${\rm d}V/{\rm d}I$)  is measured with the top voltage gate $V_{tg}=1.05 $ V, exhibiting the Fraunhofer pattern. (b) The nearly flat distribution of the supercurrent density indicates the bulk-dominated regime. (c) Differential resistance ${\rm d}V/{\rm d}I$ is measured with the top voltage gate $V_{tg}=-0.425 $ V, showing a ``SQUID-like" sinusoidal pattern. (d) The sharp peaks of the supercurrent density indicate the edge-dominated regime.  {\color{black}Reprinted} from Ref.~\citenum{QSHJJ1}. }
    \label{fig:fig2}
\end{figure*}

On the other hand, when the chemical potential lies within the bulk gap, an incident electron from one chiral edge is transformed into an outgoing hole along the opposite edge after encountering the interface between the SC and central region through the crossed or nonlocal Andreev reflections (CARs) in both QH-based and QAH-based JJs \cite{CAR1,CAR2,f2,f4,f8}. Similarly,
the hole is then reflected as an electron along the other interface. The entire process forms a loop, as depicted in Fig.~\ref{fig:fig1}(f), which is analogous to the Aharonov-Bohm (AB) oscillations \cite{AB1}. In AB oscillations, electrons traveling along different paths enclose a magnetic flux $\Phi$, causing resistance to oscillate with the magnetic flux exhibiting a periodicity of $2\Phi_0$ \cite{AB2}. Consequently, the critical Josephson current $I_c$ in response to the external magnetic flux $\Phi$ has two distinct features {\color{black}\cite{f1,f2,f3,f31,f4,f5,f6,f7,f8,f9}}: (a) The periodicity of the quantum interference pattern is $2\Phi_0$. (b) The minima of the oscillation is non-zero.

In this review, we focus on the quantum interference patterns in response to the external magnetic field in QSH-based, QH-based, and QAH-based JJs related to the different Andreev reflections. 
We begin by outlining the experimental and theoretical progress in QSH-based JJs in Sec.~\ref{subsection2.1} and  Sec.~\ref{subsection2.2}, respectively. Then, we discuss the experimental and theoretical research in QH-based JJs in Sec.~\ref{subsection3.1} and  Sec.~\ref{subsection3.2}, respectively. Following this, we introduce the theoretical developments in QAH-based JJs in Sec.~\ref{section4}. Finally, we conclude the review with a discussion on the possible future directions in Sec.~\ref{section5}.

\section{ QSH-based JJs}\label{section2}

\subsection{Experimental progress }\label{subsection2.1}

%The HgTe/HgCdTe quantum wells undergo a transition from a normal insulator to a QSH insulator (QSHI) when the well width $d>6.3$ nm.
In 2014, the observation of helical edge supercurrent was reported in a QSH-based Josephson junction utilizing a  HgTe/HgCdTe quantum well contacted by two titanium/aluminum superconducting contacts \cite{QSHJJ1}. 
%When $d<6.3$ nm, the system operates as a conventional non-topological JJ. In this regime, the bulk carriers are uniform distributed in the junction, and the critical Josephson current exhibit the typical Fraunhofer pattern in response to the external magnetic field $B$.  However, when $d>6.3$ nm, the system transitions to a topological JJ. 
The junction's central region comprised a 7.5-nm-wide quantum well, with each electrode extending over 1 $\mu$m in length  and 4 $\mu$m in width ($W$), separated by a distance ($L$) of 800 nm.  A top gate voltage $V_{tg}$ was applied to tune the carrier distribution within the topological JJ. {\color{black} When $V_{tg}=1.05$ V}, the critical Josephson current $I_c(B)$ is characterized by a Fraunhofer pattern as shown in Fig.~\ref{fig:fig2}(a).  The current density profile depicted in Fig.~\ref{fig:fig2}(b), extracted by the Dynes-Fulton procedure \cite{QSHJJ12}, reveals a nearly flat distribution, indicating the bulk-dominated regime. Conversely, {\color{black} when $V_{tg}=-0.425$ V}, the interference pattern of $I_c(B)$ approximated a sinusoidal oscillation (see Fig.~\ref{fig:fig2}(c)). This ``SQUID-like" interference exhibited a periodicity of $\Phi_0$, where $\Phi = BS$ represents the effective magnetic flux, and $S$ denotes the effective flux focusing area. Considering the Meissner effect, the effective area corresponding to this period includes the area of the HgTe region plus half the area of the superconducting electrodes on each side. The sharp peaks in the current density at the QSHE/SC interface show that the current is confined to the edge (see Fig.~\ref{fig:fig2}(d)). 
 
\begin{figure*}[ht!]
    \centering
    \includegraphics[width=1\linewidth]{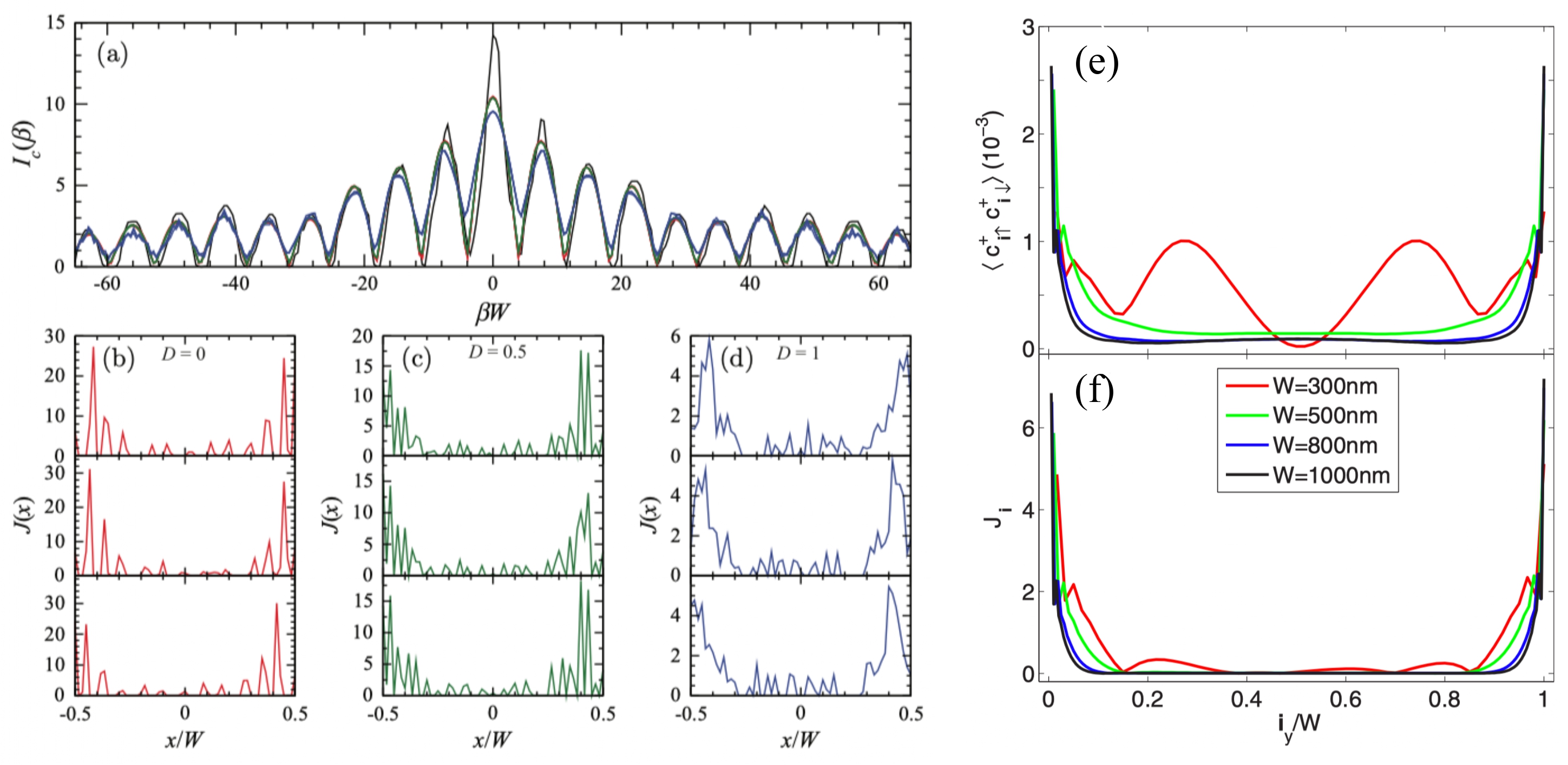}
    \caption{(a) The black line shows the quantum interference patterns obtained from the experimental data in Fig.~\ref{fig:fig2}(c). The red, green, and blue lines correspond to the quantum interference patterns obtained from the current profiles below, found with the numerical simulation, for the transparencies $D=0$, $D=0.5$, and $D=1$, respectively. In (b)-(d), three slightly different current profiles for each value of the transparencies $D=0$, $D=0.5$, and $D=1$, respectively. Subplots (a)-(d) are {\color{black}reprinted} from Ref.~\citenum{QSHJJ4}.  Distribution of (e) Cooper pair and (f) current density profile along the cross section of the central TI region. Subplots (e)-(f) are {\color{black}reprinted} from Ref.~\citenum{QSHJJ5}.  }
    \label{fig:fig3}
\end{figure*}

Additional evidence for helical edge supercurrents has been experimentally observed in a Ti/Al–InAs/GaSb–Ti/Al JJ \cite{QSHJJ3}. The junction features contact widths ($W$) of 3.9 $\mu$m and a separation ($L$) of 400 nm. A Ti/Au top gate ($V_{tg}$) and an $\rm{n}^{+}$ GaAs substrate, functioning as a back gate ($V_{bg}$), are utilized to tune the carrier distribution.  {\color{black}The authors explore} the transition from bulk-dominated to edge-dominated supercurrent, induced by gate tuning. When the Fermi energy resides in the conduction band, a Fraunhofer pattern is observed, with the corresponding current density profile indicating bulk-dominated transport.  Conversely, when the Fermi energy approaches the charge neutrality point, a ``SQUID-like" sinusoidal interference pattern emerges, characterized by a current density profile peaking at the edges. Besides, the authors also report an {\color{black} even-odd} interference pattern with a periodicity of $2\Phi_0$.  Several mechanisms potentially responsible for this even-odd pattern are discussed in Sec.~\ref{subsection2.2}.

%The differential resistance ($dV/dI$) is measured, and the corresponding supercurrent density is analyzed in three regions:(e)-(f) in the n-region  ($V_{tg}=4.8$ V and $V_{bg}=0.2$ V) and (g)-(h) at the charge neutrality point  ($V_{tg}=-0.3$ V and $V_{bg}=-0.4$ V). (i)-(j) An even-odd interference pattern at $V_{tg}=5.5$ V and $V_{bg}=-0.8$ V. Subplots (e)-(j) are adapted from Ref.~\citenum{QSHJJ3}.

All experimental studies referenced in this subsection \cite{QSHJJ1,QSHJJ3} verify the existence of edge supercurrents by meeting two criteria: a ``SQUID-like" interference pattern with a periodicity of $\Phi_0$ and an edge current density profile extracted using the Dynes-Fulton procedure \cite{QSHJJ12}. {\color{black} However, several issues merit further attention, as discussed in detail in Sec.~\ref{subsection2.2}. Notably, Ref.~\cite{QSHJJ1} reports residual resistance contributed by the bulk carriers despite the current density profile confirming edge supercurrents. Specifically, when the sample width was $W = 4$ $\mu$m, the residual resistance from bulk states was approximately $h/6e^2$, yet the quantum interference pattern still exhibited a ``SQUID-like" oscillation. Furthermore}, a quantitative deconvolution of the observed $\Phi_0$-periodic interference pattern using the Dynes-Fulton procedure reveals superconducting edge channels with widths of 180–408 nm in the HgTe/HgCdTe system \cite{QSHJJ1} and 260 nm in InAs/GaSb system \cite{QSHJJ3}. However, the quantitative analysis is not particularly convincing due to the several simplifying assumptions inherent in the Dynes-Fulton procedure \cite{QSHJJ4}.

\subsection{Theoretical progress }\label{subsection2.2}

Following the Dynes-Fulton procedure, the critical Josephson current $I_c(B)$ is obtained by the Fourier transform of the current density profile, which is used to the main experimental results discussed in Sec.~\ref{subsection2.1}. However, as noted in Ref.~\cite{QSHJJ4}, the Dynes-Fulton procedure involves several simplifying assumptions: (a) a sinusoidal current-phase relation associated with a short-junction assumption and low contact transparency ($D \ll 1 $); (b) the effective area is assumed to be independent of experimental variables such as gate voltage and external magnetic field; (3) nonuniformity and inhomogeneity of the current density along the current flow direction are ignored; and (4) the current density proﬁle extracted from $I_c(B)$ is not unique. Therefore, the authors extract the current density profile using the interference patterns data reported in Refs.~\cite{QSHJJ1,QSHJJ3} by optimizing the Dynes-Fulton procedure in three aspects: (a) allowing a nonsinusoidal current-phase relation and nonvanishing transparency ($0 \leq D \leq 1 $); (b) lifting the symmetry constraint; and (c) retaining only the solutions where currents flow inside the samples. As shown in Figs.~\ref{fig:fig3}(a)-(d), the authors conclude that their results qualitatively agree with the experimental findings, particularly regarding the peaks of the current density profile located at the edges. However, they also confirm the inadequacy of the Dynes-Fulton procedure for quantitatively extracting the current density profile due to the non-uniqueness of the current distribution extraction as demonstrated for each value of the transparencies $D=0$, $D=0.5$, and $D=1$, respectively (see Figs.~\ref{fig:fig3}(c)-(d)).  Therefore, the authors address two issues related to  the residual resistance and the widths of edge in Refs.~\cite{QSHJJ1,QSHJJ3} by the analysis of the Dynes-Fulton procedure.

However, Song et al. \cite{QSHJJ5} offer a different perspective on the residual resistance. First, they numerically calculated the Cooper pair distribution, finding contributions from both edge and bulk states to the proximity effect in a superconductor. Second, they analyzed the Cooper pair distribution and current density profile as functions of the sample width ($W$). To compare their results with those of Ref.~\cite{QSHJJ1}, they fixed the resistance at approximately $h/8e^2$ by adjusting the Fermi energy, regardless of sample width changes.  As shown in Fig.~\ref{fig:fig3}(e) and (f), both the distribution of Cooper pair and current density profile at the sample center decreases gradually with increasing sample width. Notably, when the sample width reaches $W = 1$ $\mu$m, the contribution of the bulk states to Cooper pair or supercurrent becomes negligible. Conversely, the contribution of the edge states remains steady as the sample width varies from $W = 0.3$ $\mu$m to $W = 1$ $\mu$m. Therefore, only the edge supercurrent is detectable experimentally when the sample width is $W = 4$ $\mu$m, and the residual resistance from bulk states is approximately $h/6e^2$, as reported in Ref.~\cite{QSHJJ1}.

Baxevanis et al. \cite{QSHJJ6} propose a mechanism for the even-odd quantum interference pattern observed in the QSH-based JJ. Following Ref.~\cite{QSHJJ7}, they consider an extended JJ, characterized by nonzero charge transfer between the SC and QSH insulator (QSHI) region, which locally elevates the Fermi level into the conduction band of the QSHI region.  Consequently, two narrow gapless channels may remain at some penetration length from the QSHI/SC interfaces. These channels are nonhelical and can transfer electrons (holes) between top and bottom helical edge states with either spin. As a result, both LARs and CARs can occur, with their competition giving rise to the even-odd quantum interference pattern. Tkachov et al. \cite{QSHJJ8} propose a different mechanism for the even-odd quantum interference pattern in long QSH-based JJ, where the length of the junction $L$ is much larger than the superconducting coherence length $\xi_0$. The magnetic field has a twofold effect: (a) inducing oscillations of the Josephson current due to the extra phase difference $\pm \pi \Phi/\Phi_0$; and (b) generating a finite Cooper-pair (condensate) momentum along the top and bottom edges with opposite signs, referred to as Doppler-shifted ARs. At low temperature, Doppler-shifted ARs become significant, leading to different interference patterns in the Josephson current from the top and bottom edges.  The total Josephson current, being the sum of contributions from both edges, thus exhibits an even-odd pattern due to the Doppler-shifted ARs.

\begin{figure}[ht!]
    \centering
    \includegraphics[width=3.00in]{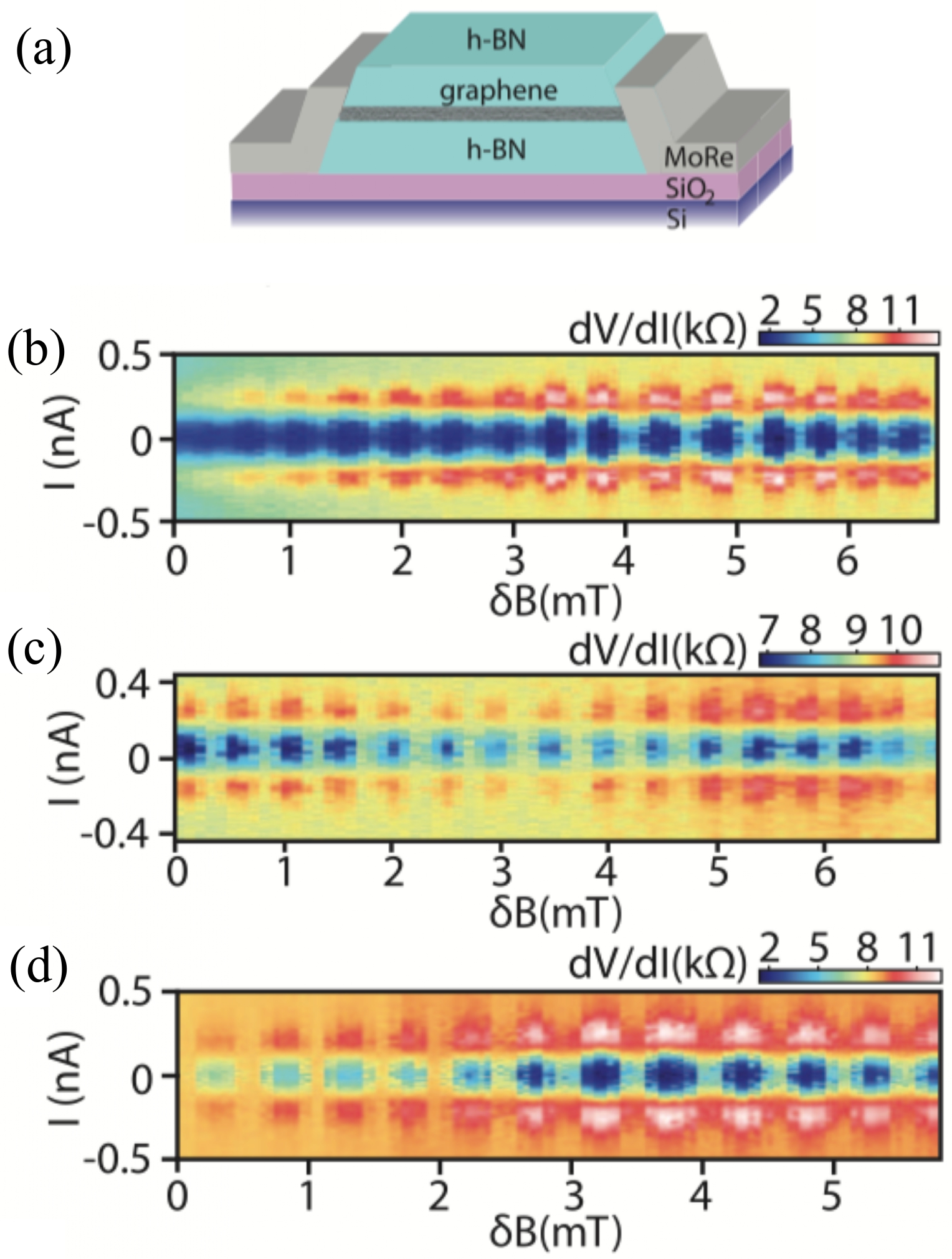}
    \caption{(a) Schematic diagram of the JJs made of graphene encapsulated in hBN and contacted by superconducting MoRe electrodes. (b-d) Differential resistance ${\rm d}V/{\rm d}I$ is measured as a function of bias current $I$ and incremental magnetic field $\delta B$ near $B=1$ T at 40 mK. Three panels correspond to the superconducting  pockets situated at fixed back-gate voltages of (b) -5.1 V, (c) -2.2 V, and (d) -2.6 V.   Adapted from Ref.~\citenum{QHJJ1}.  }
    \label{fig:fig4}
\end{figure}

\section {QH-based JJs}\label{section3}
\subsection{Experimental progress }\label{subsection3.1}

A QH-based JJ is formed by attaching superconducting leads to a sample exhibiting  QHE. The interplay between the SC and the quantum Hall state remains an experimental challenge.
This challenge is primarily attributed to the necessity of  a strong perpendicular magnetic field ($B$) to introduce the QHE within a two-dimensional electron gas. While a strong magnetic field is advantageous for the formation of Landau levels, thus rendering the bulk states insulating,  it also impedes the maintenance of superconductivity. Here, we discuss the characteristic energy and length scales to observe supercurrent carried by chiral edge states experimentally in such a JJ. The coherence length in the superconduting regions is defined as $\xi_0=\hbar v_F/\Delta_0$, where $\hbar$ is the Planck constant, $v_F$ is the Fermi velocity, and $\Delta_0$ is the superconducting pair potential. Given the potential for a high magnetic field $B$ to disrupt the SC,  it is necessary for $\xi_0$ to be small   compared to the magnetic length $l_m=\sqrt{\hbar/eB}$. The magnetic length $l_m$ provides a measure of the spatial scale over which electron wave functions spread in a magnetic field. Ensuring $\xi_0 < l_m$ guarantees that the magnetic field $B$ does not exceed the upper critical field of the SCs, beyond which superconductivity is lost. On the other hand, 
to maintain the quantum regime of the QHE, it is required that the diameter of the cyclotron orbit, calculated as $r_c = \hbar k_F/eB$ (where $k_F$ is the Fermi wavevector), is much smaller than the dimensions of the device. This requirement, expressed as $2r_c \leq \text{min}(L,W)$, ensuring the sample is in the QHE regime, not the semiclassical one. In this quantum regime, the bulk states are gapped due to the formation of Laudau levels so that the current only propagates along the edge.

The first experimental observation of supercurrent in QH-based JJs was reported by Ref.~\cite{QHJJ1} in 2016.  As illustrated in Fig.~\ref{fig:fig4}(a), these JJs consist of graphene encapsulated in hexagonal boron nitride (hBN), with molybdenum-rhenium (MoRe) serving as superconducting contacts. MoRe, a type II superconductor, features an upper critical field of $H_{c2} = 8$ T and a critical temperature ($T_c$) of approximately 8 K.  The junction has a length of  $L=0.3$ $\mu$m and a width of $W=2.4$ $\mu$m. Transport measurements conducted in a four-terminal configuration, with a back-gate voltage $V_{bg}$ used to control the carrier density in the graphene. Differential resistances, as shown in Figs.~\ref{fig:fig4}(b)-(d), reveal distinct interference patterns with a period of 0.5 mT. {\color{black}To confirm the presence of supercurrent, the authors applied a direct current bias (DC) of $I_{\rm{DC}}= 3 $ nA to the junctions, sufficient to suppress superconductivity. The disappearance of the periodic oscillation under this condition confirmed the existence of supercurrent. Furthermore, to verify that the current is carried by chiral edge states, the authors analyzed the periodic oscillations in terms of current distribution. At very low magnetic fields ($B < 10$ mT), the Fraunhofer pattern suggests that the current flows uniformly through the QH region. However, as the magnetic field increases to tens of milliteslas and beyond, the irregular pattern observed in the differential resistance indicates a non-uniform current distribution. At $B = 1$ T, the gapped bulk states restrict the current flow to the chiral edge states along the boundary, implying that the supercurrent is predominantly carried by these chiral edge states.}

As previously mentioned in Sec.~\ref{section1},  theoretical works  predict a $2\Phi_0$-periodic oscillation resulting from CARs in chiral JJs. To enable comparison with theoretical expectations, it is necessary to convert the period $B=0.5$ mT  into magnetic flux units of $\Phi_0$. Magnetic flux is defined as $\Phi=BS$, where $S$ represents the effective flux focusing area. This gives rise to a significant concern regarding the accurate estimation of the effective area. Considering the penetration depth of the magnetic field inside MoRe, the effective area is determined by the smaller of  $W^2/2$  or $W \times (C_1/2+C_2/2)$,  where $C_{1,2}$  represent the widths of the electrodes. Consequently, a $\Phi_0$-periodic oscillation diverging from theoretical predictions is observed, suggesting a possible ``SQUID-like" mechanism whereby the supercurrent may propagate independently along each edge of the QH region, rather than encircling the entire perimeter of the junction.

To investigate further, the same research group conducted a series of experiments on graphene-based JJs as shown in Fig.~\ref{fig:fig4}(a), varying their lengths, widths, and shapes \cite{QHJJ2,QHJJ3,QHJJ4}, and observed  $\Phi_0$-periodic oscillations in all cases. In particular, the study reported in Ref.~\cite{QHJJ2} reveals that supercurrent is suppressed as the junction length ($L$) increases but shows no clear dependence on junction width ($W$), which may support the hypothesis of a ``SQUID-like" behavior. {\color{black}  Furthermore, the authors designed an extended junction, in which one edge of the graphene extends beyond the superconducting contacts. } However, when comparing extended junctions to standard ones, supercurrent is efficiently suppressed in the extended junctions, suggesting the presence of  a perimeter-encircling mechanism.  In subsequent research detailed in Ref.~\cite{QHJJ3}, the experiment introduces thinly etched trenches perpendicular to the superconducting contacts at the center of the devices, with widths approximately 30 nm. Resistance measurements show that these trenches divide the central region into two parallel QH segments, each characterized by quantized Hall plateaus. This configuration is expected to double the periodicity of the interference pattern, as the encircling area for each segment is halved. Contrary to expectations, however, a $\Phi_0$ periodicity is still observed despite the presence of trenches, leaving the underlying mechanism unresolved. In Ref.~\cite{QHJJ4}, the experiment manipulates the carrier density of edge states by employing independently tunable side gates on both edges.  Supercurrents, carried by QH edge states induced by these side gates, flow independently along each edge of the device and can be individually controlled via the corresponding gates. Here, the supercurrents continue to exhibit a $\Phi_0$-periodic oscillation.  It should be noted that, the widths of the superconducting contacts in these studies are typically on the order of microns, significantly larger than the coherence length of  MoRe ($\xi_0 \leq 10$ nm). The discrepancy of the periodicity of the interference pattern between theoretical predictions and experimental observations will be discussed later.

\begin{figure}[ht!]
    \centering
    \includegraphics[width=3.00in]{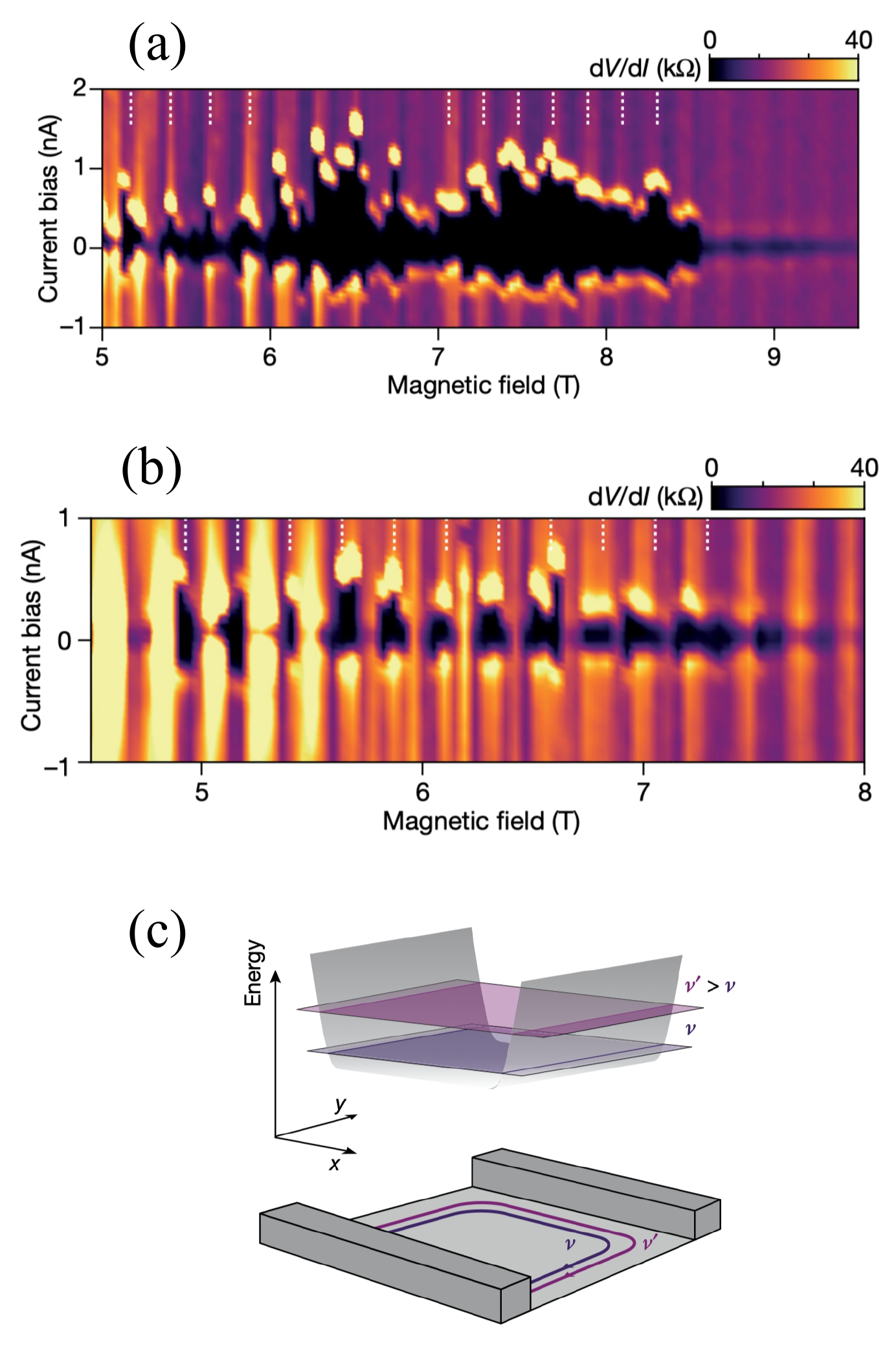}
    \caption{Differential resistance ${\rm d}V/{\rm d}I$ is measured as a function of bias current $I$ and magnetic field $\delta B$ near $B=8$ T at 0.01K at constant filling factors (a) $\nu = 2.2$ and (b) $\nu = 1.9$.  (c) Schematic diagram of Landau levels and edge states for two different filling factors at a constant magnetic field. Adapted from Ref.~\citenum{QHJJ6}.  }
    \label{fig:fig5}
\end{figure}

In 2023, an experimental study first reported the observation of a $2\Phi_0$-periodic oscillation in graphene-based JJs, as shown in Fig.~\ref{fig:fig5} (a) and (b) \cite{QHJJ6}. Compared to the experiments described in Ref.~\cite{QHJJ1}, several improvements facilitated the successful observation of this $2\Phi_0$-periodic oscillation. First, the superconducting contacts were substituted with MoGe, with a higher upper critical field ($H_{c2} \approx 12.5$ T) compared to MoRe. This alteration enabled the detection of supercurrent at elevated magnetic fields (8 T) and at a reduced filling factor ($\nu=2$).  Second, the junction dimensions were optimized, with a length of 140 nm and a width of 180 nm.  These smaller dimensions are critical as theoretical predictions suggest that the supercurrent amplitude is inversely proportional to the junction's perimeter, and long graphene/SC   interfaces can undermine coherence (this will be discussed in details later in Sec.~\ref{subsection3.2}). Third, the observation of the supercurrent at a constant filling factor, rather than a fixed back-gate voltage, is crucial.  When using a fixed back-gate voltage, the filling factor decreases as the magnetic field (B) increases. This leads to an inward displacement of the QH edge channels relative to the physical edge of the graphene, resulting in a decrease in the effective area (see Fig.~\ref{fig:fig5}(c)). Such variation in the area might render the periodicity unobservable. However, by maintaining a constant filling factor, the edge channels remain at nearly B-independent spatial positions, ensuring that the effective area remains nearly constant. Finally, a correction of area by the magnetic length $l_m$, denoted as $(L-2l_m) \times (W-2l_m)$ is considered, due to the displacement of $l_m$ relative to the physical edge of the graphene.

\subsection{Theoretical progress}\label{subsection3.2}

\begin{figure*}[ht!]
    \centering
    \includegraphics[width=1\linewidth]{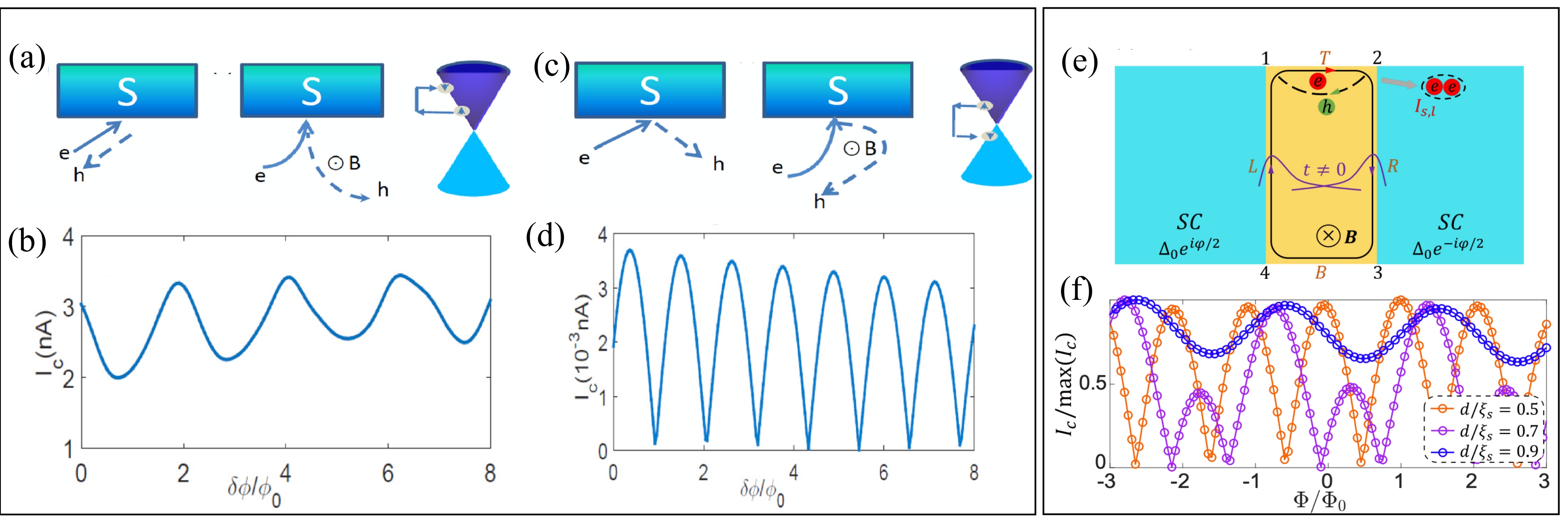}
    \caption{(a) Schematic diagram for the UAR  in the real space in the absence of the magnetic field (left panel), and in the presence of magnetic field (middle panel). Right panel: Schematic diagram for the UAR in the energy space. Here the incident electrons and reflected holes are both from the conduction band. The intraband ARs occur if $E_F>\Delta$. (b) The critical Josephson current $I_c$ versus the total  magnetic flux $\delta \phi$ in the UAR regime. (c) Schematic diagram for the SAR  in the real space in the absence of the magnetic field (left panel), and in the presence of magnetic field (middle panel). Right panel: Schematic diagram for the SAR in the energy space. Here the incident electrons and reflected holes are from the conduction and valence band, respectively. The interband ARs occur if $E_F<\Delta_0$. (d) The critical Josephson current $I_c$ versus the total  magnetic flux $\Phi$ in the SAR regime. Subplots (a)-(d) are adapted from Ref.~\citenum{f5}. (e) LARs in short junctions. In such a regime, the wavefunctions (illustrated as purple lines) of the left edge state (L) and the right edge state (R) overlap. Thus, the supercurrent is mainly mediated by the LARs. (f) The crossover from $2\Phi_0$-periodic oscillations to $\Phi_0$-periodic oscillations with the decrease of junction length $L$. Subplots (e)-(f) are adapted from Ref.~\citenum{f8}.}
    \label{fig:fig6}
\end{figure*}

First, we discuss the properties of the Josephson current. In the conventional JJs, an electron from the normal region is retroreflected as a hole after approaching the interfaces between the SC and normal metal, accompanied by the appearance of a cooper pair within the superconducting region. This process, known as Andreev retroreflection or LAR \cite{AR}, leads to the generation of the Josephson current.  When an electron traverses between SCs at the Fermi velocity $v_F$ over a separation $L$, the time taken for a complete round-trip is given by $T=2L/v_F$. Consequently, the Josephson current $I$, defined as the charge passing through a surface per unit time, is represented as $I=ev_F/L$. Nevertheless, in realistic samples, the probability of Andreev reflection is often less than unity \cite{AR1,AR2}, and the normal region may exhibit diffusive behavior. As a result, the current typically remains below the threshold $ev_F/L$. However, LAR is forbidden due to the chirality of edge states. In this regime, an electron from the top edge  would undergo the CAR, and transforming into a hole  after hitting the QH/SC interface to form a loop. Thus, the Josephson current is irrelevant to the position of the superconducting leads and depends on the perimeter size of the junction, denoted as  $\mathcal{L}=2(L+W)$ \cite{f1,f2,f3,f4}. The Josephson current $I$ is inversely proportional to $\mathcal{L}$ instead of $L$. 

On the other hand, the critical Josephson current $I_c$, carried by chiral edge states, is theoretically distinguished by  two distinct characteristics in response to the external magnetic flux $\Phi$ \cite{f4,f5,f6,f8}: (1) The oscillation exhibits the $2\Phi_0$-periodicity. (2) The minima of $I_c$ are not at zero. However, as illustrated in Sec.~\ref{subsection3.1}, various experimental findings reveal a $\Phi_0$-periodicity, rather than the theoretically predicted $2\Phi_0$-periodicity \cite{QHJJ1,QHJJ2,QHJJ3}. Several theoretical works have been undertaken to resolve the inconsistency between theoretical predictions and experimental observations \cite{f5,f6,f8}.

In Ref.~\cite{f5}, the authors display two mechanisms of ARs within the SC-graphene-SC junctions and numerically simulate the quantum interference patterns in the presence of a strong magnetic field. In the absence of the magnetic field, except the aformentioned Andreev retroreflection or usual ARs (UARs) (see the left panel of  Fig.~\ref{fig:fig6}(a)), specular ARs (SARs) - where only the perpendicular component of the velocity change sign  (depicted in the left panel of Fig.~\ref{fig:fig6}(c)) - occur. The occurrence of these processes depends on the comparison between the Fermi energy $E_F$ and the superconducting gap $\Delta_0$. Furthermore, the two AR processes exhibit very different behaviors in the presence of a strong magnetic field. In such a regime, the trajectories of incident electrons and reflected holes are bent, the Laudau levels occur, and the edge states merge near the boundary. When $E_F > \Delta_0$, UAR takes place, and the reflected holes are bent along the same directions as the incident electrons (see the middle panel of  Fig.~\ref{fig:fig6}(a)). Consequently, the holes can only propagate along the graphene/SC interface and then along the bottom edge, forming crossed Andreev pairs with the electrons on the top edge.  This process would lead to the $2\Phi_0$-periodic oscillation (see Fig.~\ref{fig:fig6}(b)). However, the UAR regime does not account for the 
$\Phi_0$-periodic oscillations observed in experiments.  When $E_F < \Delta_0$, SAR occurs, and the reflected holes are bent along the reverse directions of the incident electrons (see the middle panel of  Fig.~\ref{fig:fig6}(c)). In this scenario,  the incident electrons and the reflected holes form local Andreev pairs on a singe edge, leading to the $\Phi_0$-periodic oscillation (see Fig.~\ref{fig:fig6}(d)). Nevertheless, the critical current observed in Ref.~\cite{QHJJ1} is on the order of nA,  whereas that predicted for SAR regime is about $10^{-3}$ nA.  Therefore, the authors argue that the experimental results cannot be adequately described by either of the mechanisms discussed above. 

\begin{figure*}[ht!]
    \centering
    \includegraphics[width=1\linewidth]{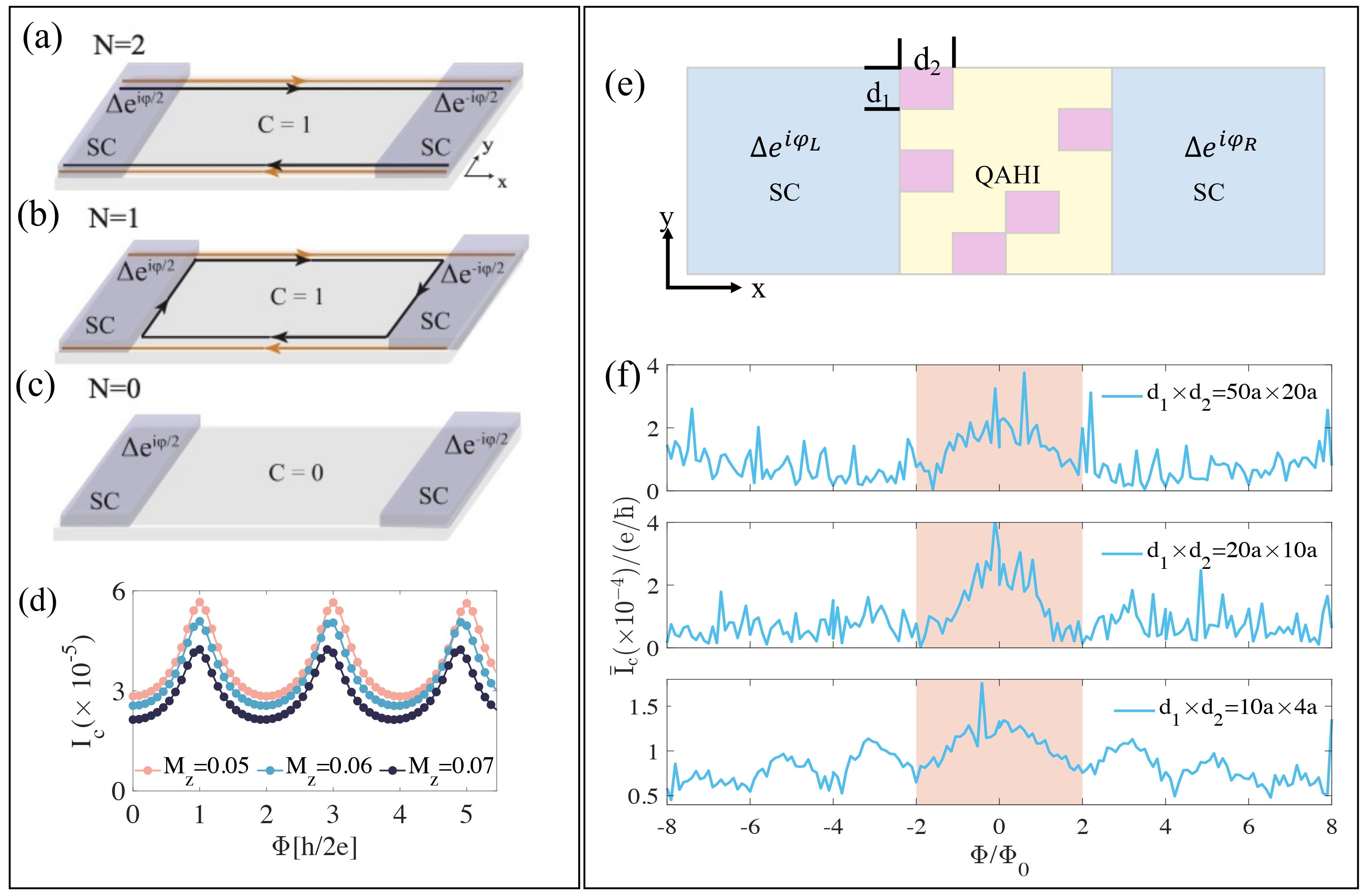}
    \caption{ Schematic plot of QAH-based Josephson junctions, consisting of two superconducting electrodes connected by a central QAHI region. The superconducting electrodes is formed when the QAHI is in proximity to an $s$-wave SCs. The QAHI/SC heterostructures can exhibit three different phases: (a) a $\mathcal{N}=2$ TSC,  (b) a $\mathcal{N}=1$ TSC, and (c) a $\mathcal{N}=0$ NSC. There are no Andreev bound states in the $\mathcal{N}=2$ and $\mathcal{N}=0$ phase. For the $\mathcal{N}=1$ phase, the edge supercurrent  is carried by a chiral Majorana mode (black loop in (b)). (d) The critical Josephson current $I_c$ oscillates with the external magnetic flux $\Phi$ characterized by a periodicity of $2\Phi_0$ in the $\mathcal{N}=1$ phase. Subplots (a)-(d) are adapted from Ref.~\citenum{f7}. (e) Schematic plot of a QAH-based JJ with a random magnetic domain structure (pink rectangle). (f) The Fraunhofer-like patterns with the central lobes marked with orange shading. Subplots (e)-(f) are adapted from Ref.~\citenum{f9}.  }
    \label{fig:fig7}
\end{figure*}

In Ref.~\cite{f6}, the authors consider the junctions holding the spin degenerate chiral quantum Hall edge in the central region sandwiched two $s$-wave SCs. When coupling with the SC, a gap is induced in the QH boundary spectrum at the interface. Consequently, proximity induced edge velocity renormalization along the SC leads and  transparency of the QH/SC interface is pivotal in obtaining the magnitude of the supercurrent.  The authors report the quantitative agreement in the magnitude of the supercurrent with the experimental results in Ref.~\cite{QHJJ1}. Besides, in the high temperature limit, the supercurrent exhibits an exponential dependency on the width of the superconducting leads  ($W$). This finding suggests that in experimental measurements, a smaller $W$ is beneficial to the observation of supercurrent. However, the authors acknowledge their inability to explain the $\Phi_0$-periodic oscillation observed experimentally.  In Ref.~\cite{f8}, the authors investigate crossover from $2\Phi_0$-periodic oscillations to $\Phi_0$-periodic oscillations in the chiral JJs with the decrease of junction length $L$ (see Fig.~\ref{fig:fig6}(f)). In the long junction limit ($L>\xi_0$), CARs dominate the transport of the system, resulting in the $2\Phi_0$-periodic oscillation. However, in short junctions ($L<\xi_0$), the wavefunctions of the left edge state (L) and the right edge state (R) may overlap as illustrated in Fig.~\ref{fig:fig6}(e). In such a regime, LARs would occur along a single edge.  When the length of junction is decreased to a critical value, LARs would dominate the transport, and leads to the $\Phi_0$-periodic oscillation. Moreover, the quantum interference patterns also exhibit a crossover from $2\Phi_0$-periodic oscillations to $\Phi_0$-periodic oscillations as the junction width $W$ increases  in short junctions. Experimental findings from Ref.~\cite{QHJJ6} reveal a $2\Phi_0$-periodic oscillation for $W=180$ nm, while those from Ref.~\cite{QHJJ1} depict a $\Phi_0$-periodic oscillation for $W=2.4$ $\rm{\mu}m$. Hence, the authors propose that the dependencies of oscillation periods on junction width $W$ can apply to the experimental results.\\

\section{QAH-based JJs}\label{section4}

Although edge  supercurrents in QH-based JJs have been experimentally verified, the 
 necessity of introducing a strong magnetic field poses significant constraints on further applications. Instead, the QAH insulators (QAHIs) exhibit chiral edge states without requiring a magnetic field.  This review focuses on the QAHI based on magnetically doped topological insulators. However, comprehensive experimental results on QAH-based JJs are still lacking due to the significant challenges posed by these experiments.  The primary challenge lies in the difficulty of achieving coexistence between QAHIs and SCs. Although magnetic doping is essential for inducing QAHIs in topological insulators, it is also detrimental to SCs, similar to the effects of strong magnetic fields in quantum Hall-based JJ experiments. Therefore, controlling the appropriate proportion of magnetic doping is crucial for the success of experiments on QAH-based JJs.

On the other hand, similar to the case of QHE, the QAHIs also posses the chiral edge states. However, unlike QHE, the edge states in a QAHI system are spin-polarized. This raises the question of whether spin-singlet edge-state supercurrent can be sustained when a QAHI is sandwiched between two $s$-wave superconducting electrodes to form a chiral JJ.  The answer is yes! Since the QAHI is realized by the magnetically doped topological insulators, the spin-orbital coupling is inherent in this system. Consequently, even when the superconducting electrodes are $s$-wave superconductors, a spin-singlet supercurrent can still propagate within this system. Thus, the critical Josephson current $I_c$ in QAH-based JJ oscillates with the magnetic flux $\Phi$, also exhibiting a periodicity of $2\Phi_0$ as the case in QH-based JJs \cite{f1,f2,f3,f31,f4,f5,f6,f8}.

The authors in Ref.~\cite{f7} study another type of chiral JJs, where a QAHI is sandwiched between two $s$-wave SCs covered on both ends (see Figs.\ref{fig:fig7}(a)-(c)) . The QAHI/SC heterostructures can exist in three distinct phases: a $\mathcal{N}=2$ TSC, a $\mathcal{N}=1$ TSC, and a $\mathcal{N}=0$ normal SC (NSC). In the $\mathcal{N}=2$ phase (see Fig.\ref{fig:fig7}(a)), a chiral fermion edge state splits into two chiral Majorana edge states, both of which inject into the right TSC. This configuration precludes Andreev bound states between the two TSCs, resulting in a negligible supercurrent. In the $\mathcal{N}=0$ phase (see Fig.\ref{fig:fig7}(c)), the absence of edge states also eliminates the supercurrent. However, in the $\mathcal{N}=1$ phase (see Fig.\ref{fig:fig7}(b)), one chiral Majorana edge mode tunnels into the right TSC while the other is reflected, allowing supercurrent propagation. Additionally, the critical Josephson current $I_c$ exhibits a periodicity of $2\Phi_0$ in response to the external magnetic flux $\Phi$ (see Fig.~\ref{fig:fig7}(d)) , which is characteristic of an ideal QAH-based JJ. 

In a real QAHI, experimental measurements reveal the coexistence of edge and bulk carriers due to magnetic dopants, even when the chemical potential resides within the bulk gap.  The authors in Ref.~\cite{f9} investigate the effects of magneitc domains induced by doping on the critical Josephson current $I_c$.  When the chemical potential moves from the bulk gap to the conduction band, the $2\Phi_0$-periodic oscillation pattern is damaged, and finally a asymmetric Fraunhofer pattern is observed. This asymmetry arises from  time-reversal symmetry breaking in QAHIs. However, magnetic domains can induce the bulk carriers even when the chemical potential is within the bulk gap. Fig.~\ref{fig:fig7}(e) illustrates the random distribution of magnetic domains (pink rectangles) in the QAHI region. This
results in the appearance of the anomalous Fraunhofer-like pattern with the main central lobes, as shown in Fig.~\ref{fig:fig7}(f).

\section{Discussions}\label{section5}

While edge supercurrents have been observed in QSH- and QH-based JJs, they remain unconfirmed in QAH-based JJs, highlighting the need for further experimental research. Demonstrating superconducting transport along helical or chiral edges is essential for realizing TSCs. However, the presence of edge supercurrents alone is insufficient to confirm the superconducting proximity effect, as these supercurrents could be carried by electrons rather than Majorana fermions.  Therefore, proving the superconducting proximity effect or identifying Majorana fermions remains a key challenge. One proposed signature of Majorana fermions is the fractional Josephson effect, characterized by a $4\pi$ periodic current-phase relation in topological JJs instead of the conventional $2\pi$ period, which leads to the absence of odd Shapiro steps \cite{FJE1,FJE2,FJE3,FJE4,FJE5,FJE6,FJE7,FJE8,FJE9,FJE10,FJE11}. In conventional JJs, Shapiro steps occur at voltages given by $V=nhf/2e$ (with $n$ integer) when a microwave drive is applied to bias the junction, where $f$ is the microwave frequency in the conventional JJs \cite{Shapiro}. However, in topological JJs, due to the $4\pi$ periodicity, only even Shapiro steps appear, with the odd steps missing.  The missing first Shapiro step has been reported by several experimental groups in topological JJs, including those based on semiconductor nanowires, topological insulators, and Dirac semimetals \cite{E1,E2,E3,E4,E5,E6,E7}. However, higher odd steps often still appear, and a more robust series of missing odd steps has only been observed in the QSHI HgTe JJ \cite{E8}. It is important to note that missing odd Shapiro steps have also been observed in topologically trivial JJs \cite{E9,E10,E11}, and various non-topological factors, such as self-heating \cite{Heating1,Heating2}, Landau-Zener transtions \cite{FJE10,FJE11,LZ1,LZ2,LZ3}, and non-constant resistance \cite{R}, can also lead to the absence of odd Shapiro steps. Therefore, distinguishing between topological and trivial origins of the missing odd Shapiro steps remains a significant challenge in confirming the presence of Majorana fermions.

\begin{acknowledgments}
The authors would like to thank  Haiwen Liu, Yang Feng, Yu-Hang Li, and Po Zhang for helpful discussions. We are grateful to the support by the National Natural Science Foundation of China (Grant No.12204053 and No.92265103)  and the Innovation Program for Quantum Science and Technology (Grant No.2021ZD0302400).

\end{acknowledgments}

\bibliographystyle{apsrev4-1}

\end{document}